# The nature of the thermal phase transition with Wilson quarks

Claude Bernard

*Department of Physics, Washington University,*

*St. Louis, MO 63130, USA*

Thomas A. DeGrand and A. Hasenfratz

*Physics Department, University of Colorado,*

*Boulder, CO 80309, USA*

Carleton DeTar

*Physics Department, University of Utah,*

*Salt Lake City, UT 84112, USA*

Steven Gottlieb

*Department of Physics, Indiana University,*

*Bloomington, IN 47405, USA*

and

*Department of Physics, Bldg. 510A,*

*Brookhaven National Laboratory*

*Upton, Long Island, NY 11973, USA*

Leo Kärkkäinen and D. Toussaint

*Department of Physics, University of Arizona,*

*Tucson, AZ 85721, USA*

R. L. Sugar

*Department of Physics, University of California,*

*Santa Barbara, CA 93106, USA*

(October 21, 1993)

## Abstract

We describe a series of simulations of high temperature QCD with two flavors of Wilson quarks aimed at clarifying the nature of the high temperature phase found in current simulations. Most of our work is with four time slices, although we include some runs with six and eight time slices for comparison. In addition to the usual thermodynamic observables we study the quark mass defined by the divergence of the axial current and the quark propagator in the Landau gauge. We find that the sharpness of the $N_t = 4$ thermal transition has a maximum around $\kappa = 0.19$ and $6/g^2 = 4.8$.

12.38.Gc, 11.15.Ha

# I. INTRODUCTION

Lattice simulations are an important source of information on the behavior of quantum chromodynamics at high temperature. Most work has been done with Kogut-Susskind quarks because of the exact remnant of chiral symmetry. Since the exact chiral symmetry of Kogut-Susskind quarks is a $U(1)$ symmetry, there is some question about how well the results reproduce the real world with its $SU(2)$ chiral symmetry. In the continuum limit the complete chiral symmetry is restored. However, in the continuum limit the results should be independent of the regularization used for the quarks. To test this it is important to study high temperature QCD with the other common form of lattice quarks, the Wilson quarks.

The first simulations of high temperature QCD with two flavors of Wilson quarks revealed a potential problem — for the values of $6/g^2$ for which most low temperature simulations were done, $4.5 \leq 6/g^2 \leq 5.7$, the high temperature transition occurs at a value of quark hopping parameter $\kappa$ for which the pion mass measured at zero temperature is quite large [1,2]. In other words, it is difficult to find a set of parameters for which the temperature is the critical temperature and the quark mass is small. Further work confirmed that the pion mass is large at the deconfinement transition for this range of $6/g^2$ [3,4]. (A recent study has concluded that for four time slices the chiral limit is reached at a very small value of $6/g^2 \approx 3.9$ [5].)

Screening masses for color singlet sources show an approach to parity doubling in the high temperature phase similar to what is seen with Kogut-Susskind quarks [2,3]. Also, measurements of the pion mass show a shallow minimum at the high temperature transition [6].

Previous simulations with Wilson fermions have located $\kappa_t$, the value of the hopping parameter at whch the high temperature crossover or phase transition occurs, as a function of $6/g^2$ for $N_t = 4$ and 6. The critical value of the hopping parameter, $\kappa_c$, for which the pion mass vanishes at zero temperature has been located with somewhat less precision [1,2,6,7,3,4]. Some measurements of hadron masses been carried out on zero temperature lattices for values of $\kappa$ and $6/g^2$ close to the $\kappa_t$ curve, allowing one to set a scale for the temperature, and to estimate $\kappa_c$ in the vacinity of the thermal transition [3,4,8].

In more recent work at $N_t = 6$ we have observed coexistence of the low and high temperature phases over long simulation times, and we have extended these observations in the present project. The change in the plaquette across the transition is much larger than for the high temperature transition with Kogut-Susskind quarks [4]. This unexplained behavior, as well as work by Hasenfratz and DeGrand on the effect of heavy quarks [9], has led us to extend our work.

This paper reports on a series of simulations with Wilson quarks at high temperature, in which we have studied a number of indicators for the nature of the phases. Using $8^3 \times 4$ lattices, we have extended earlier studies of the location of the thermal transition or crossover to $\kappa = 0.20$, 0.21 and 0.22. In the ranges of $6/g^2$ and $\kappa$ that have been studied earlier, we have done extensive simulations on $8 \times 8 \times 20 \times 4$ lattices, with additional work on $12^3 \times 6$, $12^2 \times 24 \times 6$ and $8^2 \times 20 \times 8$ lattices. For one value of $6/g^2$ we made a series of runs on $6 \times 6 \times 20 \times 4$ lattices to make sure that the effects we see are not due to the spatial size of the lattice. In this report we will concentrate on the results with four time slices. Simulations with $N_t = 6$ and 8 are still underway and will be described later.

We see a number of inexplicable effects. At large $\beta$ and small $\kappa$ the crossover from the confined phase to the high temperature phase is smooth. Beginning at $(\beta, \kappa) = (5.1, .16)$ and extending down to about $(\beta, \kappa) = (4.51, .20)$ the crossover becomes abrupt, though probably not first order. A rapid crossover is seen in the plaquette, real part of the Polyakov loop, $\bar{\psi}\psi$, the entropy, and the quark mass derived from the axial current. For $\beta < 4.5$, $\kappa > 0.20$ the transition once again becomes very smooth.

Section 2 discusses the quantities we measured, and section 3 summarizes the simulations and the results. Conclusions are in section 4.

## II. MEASURED QUANTITIES

In our simulations we have measured the expectation values of the Polyakov loop, the space-space and space-time plaquettes, the chiral condensate $\bar{\psi}\psi$, the entropy, screening masses for meson sources, the quark mass defined by the divergence of the axial current, and quark propagators in Landau gauge.

The expectation value of the Polyakov loop, $\langle P \rangle$, is simply interpreted as $\exp(-F_q/T)$, where $F_q$ is the free energy of a static test quark. With dynamical quarks $\langle P \rangle$ is always nonzero, but it increases dramatically at the high temperature transition. We also measured the space-space and space-time plaquettes, $\langle \Box_{ss} \rangle$ and $\langle \Box_{st} \rangle$. In our normalization these are equal to three on a completely ordered lattice.

The energy, pressure, entropy and $\bar{\psi}\psi$ with Wilson quarks are obtained by differentiating the partition function with respect to the temporal size, the spatial size, and the quark mass, respectively. Details are given in appendix A. We study the entropy to lowest order in $g$ and $\bar{\psi}\psi$, using the formulae:

$$\left\langle \bar{\psi}\psi \right\rangle = \frac{1}{N_s^3 N_t} \frac{4\kappa N_f}{2} \operatorname{Re} \left\langle \operatorname{Tr} \frac{1}{M^\dagger} \right\rangle \tag{1}$$

and

$$\begin{aligned}
\left\langle s_g a^4 \right\rangle &\frac{1}{aN_t} \\
&= \frac{1}{3N_s^3 N_t} \sum_x \left\langle -\frac{\partial S_g}{\partial \alpha_t} + \frac{1}{3}\frac{\partial S_g}{\partial \alpha_x} + \frac{1}{3}\frac{\partial S_g}{\partial \alpha_y} + \frac{1}{3}\frac{\partial S_g}{\partial \alpha_z} \right\rangle \\
&= \frac{1}{N_s^3 N_t} \sum_x \left\langle \left(\frac{8}{g^2} + 4(C_\tau - C_\sigma)\right)(\Box_{st} - \Box_{ss}) \right\rangle
\end{aligned} \tag{2}$$

and

$$\begin{aligned}
\left\langle s_f a^4 \right\rangle \frac{1}{aN_t} &= \epsilon_f a^4 + p_f a^4 \\
&= \frac{-2}{N_s^3 N_t} \frac{\kappa N_f}{2} \operatorname{Re} \left\langle \operatorname{Tr} \frac{1}{M^\dagger} \left(\not{D}_0 - \frac{1}{3}\sum_i \not{D}_i\right) \right\rangle
\end{aligned} \tag{3}$$

where $s_g$ and $s_f$ are the gluon and fermion entropies, respectively.

We measured screening masses for meson sources with quantum numbers of the $\pi$, $\sigma$, $\rho$ and $a_1$. These measurements are a standard hadron spectrum calculation, except that the propagation is in the z direction. We used a wall source covering the entire $z = 0$ slice of the lattice, with the gauge fixed to a spatial Coulomb gauge which maximizes the traces of the $x$, $y$ and $t$ direction links. After blocking five to ten measurements together to minimize the autocorrelations, we fit all the propagators to a single exponential using the full covariance matrix of the propagator elements.

A quark mass can be defined from the divergence of the axial current [10,11]. The basic relation is a current algebra relation

$$\nabla_\mu \cdot \langle \bar\psi \gamma_5 \psi(0) \bar\psi \gamma_5 \gamma_\mu \psi(x) \rangle = 2 m_q \langle \bar\psi \gamma_5 \psi(0) \bar\psi \gamma_5 \psi(x) \rangle \tag{4}$$

If we sum over $x,y,t$ slices, and measure distance in the $z$ direction, this becomes:

$$\frac{\partial}{\partial z} \sum_{x,y,t} \langle \bar\psi \gamma_5 \psi(0) \bar\psi \gamma_5 \gamma_3 \psi(x) \rangle$$
$$= 2 m_q \sum_{x,y,t} \langle \bar\psi \gamma_5 \psi(0) \bar\psi \gamma_5 \psi(x) \rangle \tag{5}$$

We define $PS(z)$ as the pion correlator with a point sink:

$$PS(z) = \left\langle W(0) \bar\psi \gamma_5 \psi(z) \right\rangle \tag{6}$$

and $A(z)$ as the axial current correlator:

$$A(z) = \left\langle W(0) \bar\psi \gamma_5 \gamma_z \psi(z) \right\rangle \tag{7}$$

where $W(0)$ is the wall source at $z = 0$. At long distances both $PS(z)$ and $A(z)$ will fall off as $\exp(-m_\pi z)$. Therefore we perform a simultaneous fit to the two propagators on a lattice periodic in the $z$ direction using three parameters, $C$, $m_\pi$ and $m_q$,

$$PS(z) = C \sinh(m_\pi) \left[ \exp(-m_\pi z) + \exp(m_\pi (N_z - z)) \right] \tag{8}$$

$$A(z) = C\, 2\, m_q \left[ \exp(-m_\pi z) - \exp(m_\pi (N_z - z)) \right] \tag{9}$$

The factor of $\sinh(m_\pi)$ in Eq. (8) comes from using the lattice difference $f(z+1) - f(z-1)$ for the derivative in Eq. (5). Note that $PS(z)$ is periodic in $z$ while $A(z)$ is antiperiodic. We use the pointlike axial current $\bar\psi(z) \gamma_5 \gamma_\mu \psi(z)$ rather than a point split current. These are quark masses in lattice units; to convert to continuum quark masses requres a lattice-to-continuum renormalization. See Ref. [17] for a discussion of this point.

The quark propagator in the Landau gauge was also measured. This propagator has been studied with Kogut-Susskind quarks in Ref. [12]. We chose a source constant in the $y$ direction and a $\delta$ function in $x$, $z$ and $t$ with only the real part of the first Dirac component non-zero (in the Weyl basis we use). Because of the $\delta$-function all possible momenta in $x$, $z$ and $t$ directions were excited. To distinguish among the different momenta we performed a Fourier transform of the propagator in $x$ and $t$ directions (taking into account that it has to have odd frequencies in $t$ direction). This gives the propagation of the quark in the $z$ direction as function of $k_x$ and $k_t$, i.e. the dispersion relation of the screening propagator. In order to keep the amount of generated data at a reasonable level, the propagator was saved only for on-axis momentum values of $k_x$ and $k_t$. This enabled us to measure the on-axis dispersion relation of the quark screening mass, in particular the screening mass difference of the quark and light doublers.

The form to which the spatial propagator is fitted is usually motivated by the form of the free propagator. We suppose that at large distances, each separate momentum component of the spatial propagator resembles the corresponding free quark form, but with its own renormalized quark mass, or in this case of Wilson fermions, with a renormalized $\kappa$.

In momentum space the free Wilson propagator is

$$G(k) = \frac{[1 - 2\kappa \sum_\mu \cos(p_\mu)] - i2\kappa \sum_\mu \gamma_\mu \sin(p_\mu)}{[1 - 2\kappa \sum_\mu \cos(p_\mu)]^2 + 4\kappa^2 \sum_\mu \sin^2(p_\mu)}. \tag{10}$$

With our choice of the source the first Dirac component of the propagator, $G_1$, is real:

$$G_1 = \frac{[1 - 2\kappa \sum_\mu \cos(p_\mu)]}{[1 - 2\kappa \sum_\mu \cos(p_\mu)]^2 + 4\kappa^2 \sum_\mu \sin^2(p_\mu)}. \tag{11}$$

Then, for non-zero $z$ values

$$G(p_1, p_2, z, p_0) = \sum_{k=1}^{L_z} \exp[i2\pi kz/L_z] G(p_1, p_2, k, p_0)$$
$$= L_z \left[ \frac{(1 - 6\kappa + 4\kappa A)^2 - 4\kappa(B + 1)}{8\kappa(1 - 6\kappa + 4\kappa A)^2} \right] \left[ \frac{\cosh[ma(z - L_z/2)]}{\sinh(ma)\sinh(maL_z/2)} \right], \tag{12}$$

where

$$A = \sum_{\mu=1,2,0} \sin^2[p_\mu/2], \tag{13}$$

$$B = \sum_{\mu=1,2,0} \sin^2[p_\mu], \tag{14}$$

and

$$ma = 2\sinh^{-1}\left(\sqrt{\frac{4\kappa^2 B + (1 - 8\kappa + 4\kappa A)^2}{8\kappa(1 - 6\kappa + 4\kappa A)}}\right). \tag{15}$$

At zero momentum (on a low temperature lattice where the lowest Matsubara frequency is close to zero) this relation turns into

$$ma = \ln\left[\frac{1 - 6\kappa}{2\kappa}\right]. \tag{16}$$

The mass vanishes when $\kappa \to 1/8$ as expected. Inverting this for $\kappa$ gives

$$\kappa = \frac{1}{2\exp[ma] + 6}, \tag{17}$$

which, for small masses reduces to the naive relation,

$$\kappa = \frac{1}{2ma + 8} \qquad (18)$$

that one expects looking at the terms of the Lagrangian. For large lattices the lowest doubler mass becomes

$$ma_{\text{doubler}} = \lim_{\kappa \to 1/8} \ln\left[\frac{1-2\kappa}{2\kappa}\right] = \ln[3] = 1.09861. \qquad (19)$$

For Kogut-Susskind fermions [12] the free propagator turns out to be a sum of two terms, having parts with an alternating sign in $z$-direction. For Wilson fermions, with our choice of the source, the propagator is a single exponential, or hyperbolic cosine, on a finite lattice.

Furthermore, the sign of $G_1$ at $k = 0$ changes at $\kappa_c = 1/8$. Therefore, measuring the sign of the propagator can be used as an indicator of whether $\kappa$ is effectively greater or less than $\kappa_c$.

One can infer from Eqs. 13–15 that the only effect of finite spatial lattice size is the discretisation of the momenta. For a given momentum all lattice sizes give the same value of the screening mass. For a smaller lattice, the range of allowed momenta is more restricted, of course.

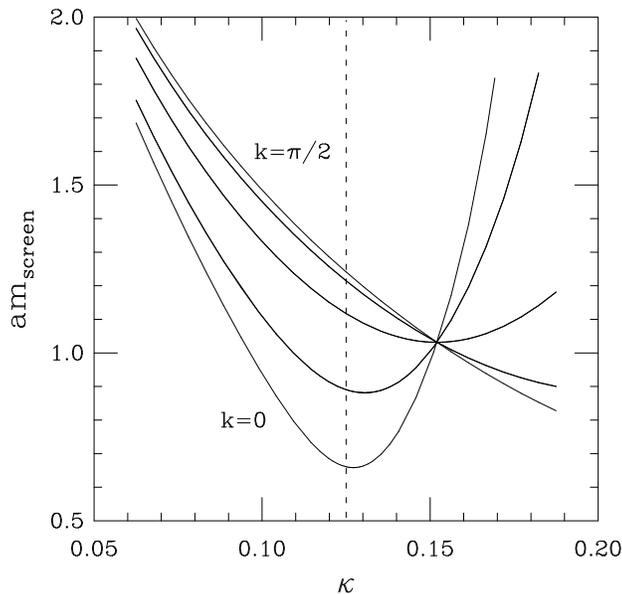

FIG. 1. The spatial screening mass at different spatial momenta for free Wilson fermions as a function of $\kappa$ for an $8^2 \times 20 \times 4$ lattice. At $\kappa_c = 1/8$ the higher masses are for higher momenta.

To be specific let us look at what happens with our lattice size: $8^2 \times 20 \times 4$. This is shown in figure 1. At $\kappa_c$ the lowest momentum screening mass is at its minimum. If one increases $\kappa$ the screening masses start to converge to a single value close to one at $\kappa = 0.152$. At this point the dispersion relation is flat.

The sign of the propagator with this source depends on the momentum. Generally, the value of $\kappa$ at which the sign changes increases with the momentum. For zero momentum it occurs at $\kappa_c$; for the smallest nonzero momentum in our lattice size it takes place at $\kappa = (2 + \sqrt{2} - \sqrt{3})/11 = 0.1374$. The amplitude for the doubler does not change sign in this kappa range.

For our lattice size, inserting the appropriate momenta to Eq. (15) one obtains the following quark screening masses at $\kappa_c = 1/8$:

$$\begin{aligned} k &= (0, 0, 0, \pi/4) & ma &= 0.6610 \\ k &= (\pi/4, 0, 0, \pi/4) & ma &= 0.8906 \\ k &= (\pi/2, 0, 0, \pi/4) & ma &= 1.1171 \\ k &= (3\pi/4, 0, 0, \pi/4) & ma &= 1.2149 \\ k &= (\pi, 0, 0, \pi/4) & ma &= 1.2411 \end{aligned} \quad (20)$$

For purely temporal momenta the free field screening mass is

$$\begin{aligned} k &= (0, 0, 0, \pm\pi/4) &, ma &= 0.6610 \\ k &= (0, 0, 0, \pm 3\pi/4) &, ma &= 1.0711 \end{aligned} \quad (21)$$

Hence, the temporal doubler's screening mass is smaller than that of the spatial doubler.

## III. SIMULATIONS AND RESULTS

| $N_t$ | $N_{x,y}$ | $6/g^2$ | $\kappa$ | traj. | ignore | dt | accept |
|---|---|---|---|---|---|---|---|
| 4 | 8 | 4.9 | 0.180 | 650 | 100 | 0.02 | 0.88 |
| 4 | 8 | 4.9 | 0.181 | 320 | 100 | 0.02 | 0.94 |
| 4 | 8 | 4.9 | 0.182 | 810 | 100 | 0.02 | 0.90 |
| 4 | 8 | 4.9 | 0.1825 | 824(c) | 100 | 0.02 | 0.86 |
| 4 | 8 | 4.9 | 0.1825 | 780(h) | 100 | 0.02 | 0.88 |
| 4 | 8 | 4.9 | 0.183 | 624(h) | 100 | 0.02 | 0.90 |
| 4 | 8 | 4.9 | 0.183 | 810(c) | 600 | 0.02 | 0.87 |
| 4 | 8 | 4.9 | 0.184 | 610 | 100 | 0.02 | 0.93 |
| 4 | 8 | 5.0 | 0.173 | 540 | 100 | 0.025 | 0.86 |
| 4 | 8 | 5.0 | 0.175 | 500 | 100 | 0.02 | 0.92 |
| 4 | 8 | 5.0 | 0.177 | 400 | 100 | 0.02 | 0.90 |
| 4 | 8 | 5.0 | 0.178 | 474 | 100 | 0.02 | 0.94 |
| 4 | 8 | 5.0 | 0.180 | 630 | 100 | 0.02 | 0.92 |
| 4 | 8 | 5.0 | 0.182 | 256 | 100 | 0.02 | 0.86 |
| 6 | 8 | 5.0 | 0.175 | 240 | 80 | 0.02 | 0.94 |
| 6 | 8 | 5.0 | 0.180 | 360 | 60 | 0.0167 | 0.87 |
| 8 | 8 | 5.0 | 0.175 | 350 | 100 | 0.0167 | 0.91 |

TABLE I. Table of runs at fixed $6/g^2$ with varying $\kappa$. "(h)" and "(c)" indicate hot and cold starts.

Simulations were run on the Intel iPSC/860 and Paragon, and the nCUBE-2 at the San Diego Supercomputer Center, on the Thinking Machines Corporation CM5 at the National

| $N_t$ | $N_{x,y}$ | $6/g^2$ | $\kappa$ | traj. | ignore | dt | accept |
|---|---|---|---|---|---|---|---|
| 4 | 8 | 5.1 | 0.165 | 1120 | 100 | 0.025 | 0.89 |
| 4 | 8 | 5.1 | 0.167 | 2660 | 100 | 0.025 | 0.89 |
| 4 | 8 | 5.1 | 0.169 | 460 | 100 | 0.025 | 0.87 |
| 4 | 8 | 5.1 | 0.170 | 700 | 100 | 0.025 | 0.90 |
| 4 | 8 | 5.1 | 0.171 | 980 | 100 | 0.025 | 0.89 |
| 4 | 8 | 5.1 | 0.172 | 3380 | 100 | 0.025 | 0.87 |
| 4 | 8 | 5.1 | 0.173 | 500 | 100 | 0.025 | 0.85 |
| 4 | 8 | 5.1 | 0.175 | 460 | 100 | 0.025 | 0.86 |
| 4 | 8 | 5.1 | 0.177 | 1800 | 100 | 0.025 | 0.88 |
| 4 | 8 | 5.1 | 0.179 | 660 | 100 | 0.025 | 0.88 |
| 4 | 6 | 5.1 | 0.169 | 860 | 200 | 0.0333 | 0.82 |
| 4 | 6 | 5.1 | 0.170 | 1000 | 100 | 0.0333 | 0.85 |
| 4 | 6 | 5.1 | 0.171 | 1360 | 100 | 0.0333 | 0.81 |
| 4 | 6 | 5.1 | 0.172 | 1120 | 100 | 0.0333 | 0.83 |
| 4 | 6 | 5.1 | 0.175 | 720 | 100 | 0.0333 | 0.81 |
| 8 | 8 | 5.1 | 0.167 | 512 | 100 | 0.025 | 0.82 |
| 8 | 8 | 5.1 | 0.173 | 279 | 100 | 0.02 | 0.87 |
| 8 | 8 | 5.1 | 0.177 | 440 | 100 | 0.02 | 0.66 |
| 4 | 8 | 5.3 | 0.155 | 2400 | 100 | 0.0333 | 0.78 |
| 4 | 8 | 5.3 | 0.157 | 660 | 100 | 0.0333 | 0.76 |
| 4 | 8 | 5.3 | 0.158 | 1239 | 100 | 0.0333 | 0.89 |
| 4 | 8 | 5.3 | 0.159 | 660 | 100 | 0.0333 | 0.89 |
| 4 | 8 | 5.3 | 0.160 | 1777 | 100 | 0.025 | 0.88 |
| 4 | 8 | 5.3 | 0.161 | 480 | 100 | 0.025 | 0.88 |
| 4 | 8 | 5.3 | 0.162 | 480 | 100 | 0.025 | 0.83 |
| 4 | 8 | 5.3 | 0.163 | 680 | 100 | 0.025 | 0.87 |
| 4 | 8 | 5.3 | 0.164 | 460 | 100 | 0.025 | 0.90 |
| 4 | 8 | 5.3 | 0.165 | 720 | 100 | 0.025 | 0.88 |
| 4 | 8 | 5.3 | 0.166 | 660 | 100 | 0.025 | 0.87 |
| 4 | 8 | 5.3 | 0.167 | 912 | 100 | 0.025 | 0.86 |
| 4 | 8 | 5.3 | 0.168 | 540 | 100 | 0.025 | 0.89 |
| $4^\dagger$ | 8 | 5.3 | 0.168 | 840 | 100 | 0.025 | 0.80 |
| 4 | 8 | 5.3 | 0.169 | 380 | 100 | 0.025 | 0.86 |
| 4 | 8 | 5.3 | 0.170 | 380 | 100 | 0.025 | 0.79 |
| 4 | 8 | 5.3 | 0.172 | 440 | 100 | 0.025 | 0.87 |
| 6 | 12 | 5.3 | 0.155 | 320 | 100 | 0.0177 | 0.88 |
| 6 | 12 | 5.3 | 0.160 | 552 | 60 | 0.0177 | 0.91 |
| 6 | 12 | 5.3 | 0.165 | 666 | 216 | 0.0177 | 0.85 |
| 6 | 12 | 5.3 | 0.166 | 1403 | 400 | 0.0177 | 0.84 |
| 6 | 12 | 5.3 | 0.167 | 760 | 302 | 0.0177 | 0.84 |
| 6 | 12 | 5.3 | 0.168 | 603 | 200 | 0.0177 | 0.85 |

TABLE II. Table of runs at fixed $6/g^2$ with varying $\kappa$. "(h)" and "(c)" indicate hot and cold starts. Most of the $N_t = 4$ runs were on $8^2 \times 20 \times 4$ lattices. The run indicated with a † at $\kappa = 0.168$, $6/g^2 = 5.3$ was done on a $8^2 \times 40 \times 4$ lattice.

| $N_t$ | $N_{x,y,z}$ | $6/g^2$ | $\kappa$ | traj. | ignore | dt | accept |
|---|---|---|---|---|---|---|---|
| 4 | 8 | 4.75 | 0.19 | 578 | 100 | 0.014286 | 0.912(13) |
| 4 | 8 | 4.755 | 0.19 | 1504 | 750 | 0.014286 | 0.960(6) |
| 4 | 8 | 4.76(c) | 0.19 | 837 | 500 | 0.014286 | 0.948(11) |
| 4 | 8 | 4.76(h) | 0.19 | 362 | 100 | 0.014286 | 0.962(12) |
| 4 | 8 | 4.32 | 0.20 | 156 | 50 | 0.02 | 0.83(4) |
| 4 | 8 | 4.36 | 0.20 | 172 | 50 | 0.02 | 0.83(3) |
| 4 | 8 | 4.40 | 0.20 | 188 | 50 | 0.02 | 0.79(3) |
| 4 | 8 | 4.44 | 0.20 | 152 | 50 | 0.02 | 0.77(4) |
| 4 | 8 | 4.48 | 0.20 | 368 | 50 | 0.02 | 0.69(3) |
| 4 | 8 | 4.50 | 0.20 | 244 | 100 | 0.014286 | 0.84(3) |
| 4 | 8 | 4.52 | 0.20 | 841 | 150 | 0.014286 | 0.773(16) |
| 4 | 8 | 4.54 | 0.20 | 566 | 150 | 0.014286 | 0.72(2) |
| 4 | 8 | 4.56 | 0.20 | 1324 | 200 | 0.014286 | 0.941(7) |
| 4 | 8 | 4.60 | 0.20 | 365 | 50 | 0.014286 | 0.937(14) |
| 4 | 8 | 4.64 | 0.20 | 244 | 50 | 0.02 | 0.959(14) |
| 4 | 8 | 4.10 | 0.21 | 74 | 50 | 0.01 | 0.79(8) |
| 4 | 8 | 4.20 | 0.21 | 267 | 50 | 0.005 | 0.90(2) |
| 4 | 8 | 4.26 | 0.21 | 586 | 100 | 0.005 | 0.85(2) |
| 4 | 8 | 4.28 | 0.21 | 478 | 100 | $0.005 \to 0.0025$ | 0.82(2) |
| 4 | 8 | 4.30 | 0.21 | 454 | 100 | 0.005 | 0.904(16) |
| 4 | 8 | 4.32 | 0.21 | 227 | 50 | 0.005 | 0.94(2) |
| 4 | 8 | 4.34 | 0.21 | 259 | 50 | 0.007143 | 0.943(15) |
| 4 | 8 | 4.36 | 0.21 | 281 | 100 | $0.0025 \to 0.007143$ | 0.960(14) |
| 4 | 8 | 4.40 | 0.21 | 197 | 50 | $0.0025 \to 0.01$ | 0.95(2) |
| 4 | 8 | 4.44 | 0.21 | 249 | 50 | $0.007143 \to 0.01$ | 0.977(9) |
| 4 | 8 | 4.50 | 0.21 | 120 | 50 | $0.05 \to 0.01$ | 0.94(3) |
| 4 | 8 | 3.80 | 0.22 | 59 | 15 | 0.001 | 0.93(4) |
| 4 | 8 | 3.90 | 0.22 | 56 | 30 | $0.002 \to 0.0004$ | 0.69(9) |
| 4 | 8 | 3.96 | 0.22 | 39 | 25 | $0.002 \to 0.0005$ | 0.36(13) |
| 4 | 8 | 4.00 | 0.22 | 119 | 80 | $0.004 \to 0.002$ | 0.50(8) |
| 4 | 8 | 4.04 | 0.22 | 161 | 50 | 0.004 | 0.71(4) |
| 4 | 8 | 4.06 | 0.22 | 119 | 40 | 0.003333 | 0.78(5) |
| 4 | 8 | 4.10 | 0.22 | 234 | 50 | 0.005 | 0.86(3) |
| 4 | 8 | 4.20 | 0.22 | 90 | 50 | 0.007143 | 0.98(3) |
| 4 | 8 | 4.30 | 0.22 | 58 | 50 | $0.005 \to 0.007143$ | 1.00(0) |
| 4 | 8 | 4.40 | 0.22 | 90 | 50 | 0.01 | 0.95(3) |
| 4 | 8 | 4.50 | 0.22 | 122 | 50 | $0.01 \to 0.02$ | 0.96(2) |

TABLE III. Table of runs at fixed $\kappa$ with varying $6/g^2$. "(h)" "(c)" indicate hot and cold starts. The acceptance rate gives the average over all runs in the sample kept for measurement, whether or not $dt$ was changing during the runs.

Center for Supercomputing Applications, and on a cluster of RS6000 workstations at the University of Utah. We used the hybrid Monte Carlo algorithm with two flavors of dynamical quarks in all our simulations [13]. The parameters of our runs are listed in tables I, II and III.

For the $8^2 \times 20 \times 4$ runs we used trajectories with a length of one unit of simulation time and made measurements after every second trajectory. The step size for these runs ranged in the normalization of Ref. [14] from 0.033 for the largest $6/g^2$ and smallest $\kappa$ to 0.02 at the other extreme. Acceptance rates for these runs range from 70% to 90%, with an average over all the runs of 87%. For computation of the fermion force in the updating and the propagators in the measurements we used the conjugate gradient algorithm with even-odd ILU preconditioning [15]. The conjugate gradient residual, defined as $\left|\tilde{M}^\dagger \tilde{M} x - b\right| / |b|$ where $\tilde{M}$ is the preconditioned matrix, $b$ is the source vector and $x$ is the solution vector, was $10^{-6}$. Runs were made at $6/g^2 = 5.3, 5.1, 5.0$ and $4.9$ with $N_t = 4$. At $6/g^2 = 5.3$ we also made a series of runs with $N_t = 6$. At $6/g^2 = 5.1$ we ran two points with $N_t = 8$ and at $6/g^2 = 5.0$ two points with $N_t = 6$. We also ran a series of simulations at $6/g^2 = 5.1$ on $6^2 \times 20 \times 4$ lattices to verify that the spatial size of the lattice was not seriously affecting our results. At $6/g^2 = 5.3$ a series of short runs on $6^3 \times 4$ lattices was made for very large $\kappa$.

For reference we show a phase diagram for the relevant range of $\kappa$ and $6/g^2$ in Fig. 2. Previous work showed that as $\kappa$ increased from 0.16 to 0.19 along the $N_t = 4$ high temperature crossover line the pion mass decreased, suggesting a closer approach to the high temperature transition in the chiral limit [3]. More recent work by Iwasaki *et al.*, beginning from the $6/g^2 = 0$ limit, suggested that a high temperature transition for zero quark mass might be found at $\kappa \approx 0.225$ [16]. We have done a series of runs on $8^3 \times 4$ lattices in which we varied $6/g^2$ at $\kappa = 0.20, 0.21$ and $0.22$ to extend the previous work. As expected, the number of conjugate gradient iterations required in the updating increases as $\kappa$ increases in this range, and the size of the possible updating time step decreases. Thus these runs have limited statistics. In Fig. 3 we show the plaquette and Polyakov loop as a function of $6/g^2$ for the various values of $\kappa$. Notice that the transition appears to be sharpest at $\kappa \approx 0.19$, becoming smoother for larger and smaller $\kappa$. Even in those cases where the transition is very abrupt, we do not see the sorts of metastability and tunneling characteristic of strong first order transitions. We do find cases where equilibration takes a long time. The worst case was in the run at $6/g^2 = 4.9$ and $\kappa = 0.1825$. In this case we have plotted two points, from hot and cold starts. These points are marked by arrows in Fig. 3. However, these two runs eventually converged to similar values, lying in between the values in the early parts of the runs. The time history of the Polyakov loop in these two runs is shown in Fig. 4.

We now examine the $8^2 \times 20 \times 4$ runs in more detail. Figure 5 shows the real part of the Polyakov loop as a function of $\kappa$ for the different values of $6/g^2$. For $6/g^2 = 5.3$ we also include values for $N_t = 6$ to show how the transition point moves as $N_t$ increases. For all of these values of $6/g^2$ we see the expected sharp increase in the Polyakov loop at a values of $\kappa$ less than $\kappa_c$, where $\kappa_c$ is the value at which the squared pion mass vanishes on a zero temperature lattice. We estimate $\kappa_c$ at these values of $6/g^2$ from published values of $\kappa_c$ in Refs. [7] and [4] and a recent measurement at $6/g^2 = 5.3$ by the HEMCGC group: $\kappa_c(5.3) = 0.16794$ [17]. From a quadratic fit to these values, shown by a line in Fig. 2, we find $\kappa_c(6/g^2) = 0.1687(2)$ at 5.3, $0.1795(4)$ at 5.1, $0.1861(12)$ at 5.0 and $0.1941(40)$ at 4.9.

Although not a physical quantity, the number of conjugate gradient iterations used in solving $\tilde{M}^\dagger \tilde{M} x = b$ indicates how singular $\tilde{M}$ is on the average. This quantity has been used as a probe of the physics in Ref. [16] In Fig. 6 we show the average number of conjugate gradient iterations used in an updating step, where a linear extrapolation of the last two time steps was used to produce a starting guess for the solution vector. For $6/g^2 = 5.3$ and $N_t = 4$ there is very little effect on the number of iterations at $\kappa_t$. As $6/g^2$ is decreased for $N_t = 4$ there is an increasingly sharp peak in the number of iterations at $\kappa_c$. Notice also the sharp peak in the $N_t = 6$ results for $6/g^2 = 5.3$.

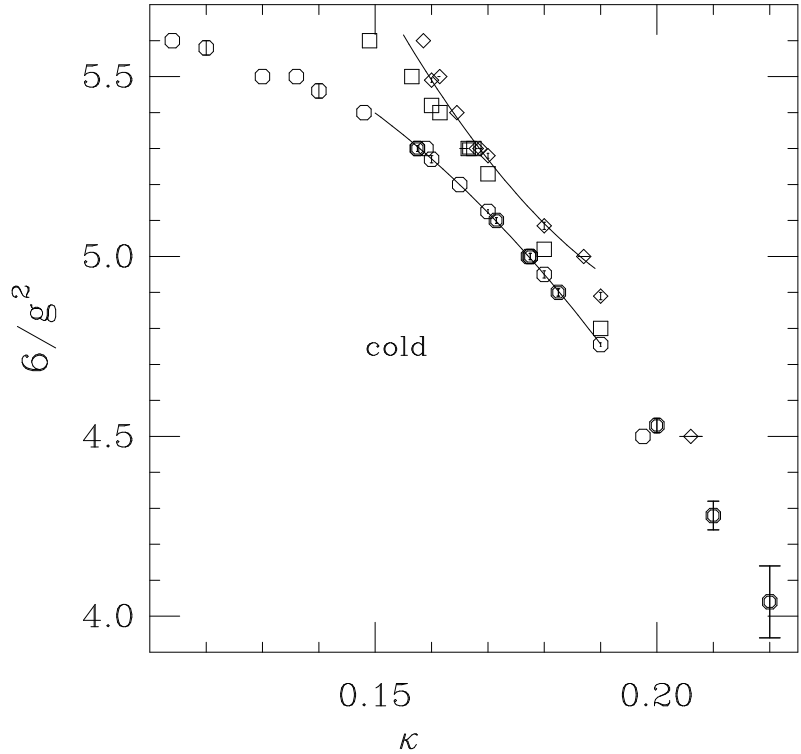

FIG. 2. Phase diagram showing estimates for the high temperature transition and $\kappa_c$. Circles represent the high temperature transition or crossover for $N_t = 4$, squares the high temperature transition for $N_t = 6$ and diamonds the zero temperature $\kappa_c$. Previous work included in this figure is from Refs. [2], [6], [7], [3] and [4]. We show error bars where they are known. For series of runs done at fixed $\kappa$ the error bars are vertical, while for series done at fixed $6/g^2$ the bars are horizontal. Points coming from this work are shown in heavier symbols. The solid lines are fits to $\kappa_t$ for $N_t = 4$ and to $\kappa_c$ used in interpolating and extrapolating.

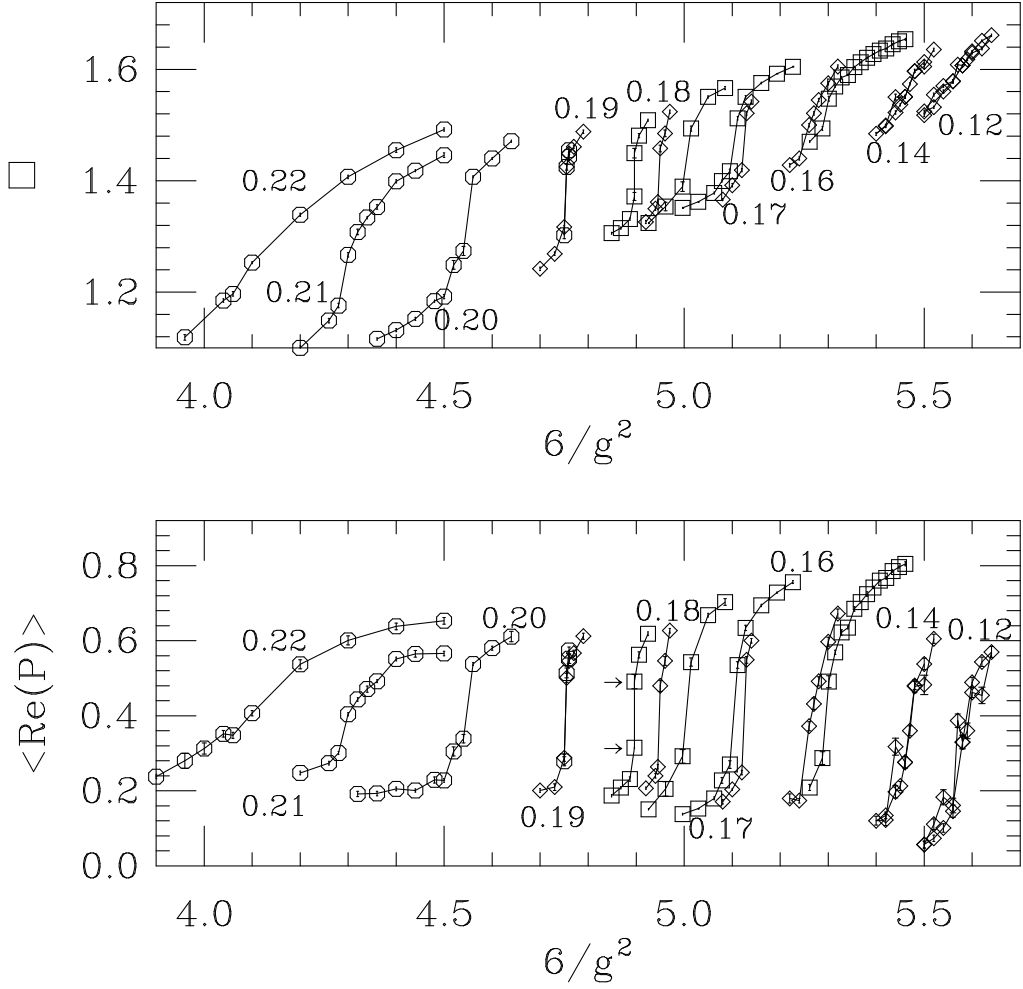

FIG. 3. The plaquette and Polyakov loop as a function of $6/g^2$ for various values of $\kappa$. The diamonds are previous results of Ref. [3] for $\kappa = 0.12, 0.14, 0.16, 0.17, 0.18$ and $0.19$. For $\kappa = 0.12$ and $0.14$ data from long runs as well as some data from short runs collected while generating hysteresis loops is shown. The octagons at $\kappa = 0.20, 0.21$ and $0.22$ are new results from $8^3 \times 4$ lattices. The squares come from runs on $8^2 \times 20 \times 4$ lattices. These runs were done at fixed values of $6/g^2$ with varying $\kappa$. They have been mapped onto this figure by fitting the $6/g_t^2$, $\kappa_t$ line (with a fit shown as a line in Fig. 2), and moving the points in the $\kappa$, $6/g^2$ plane parallel to this line. Specifically, we plot the points at $6/g^2_{effective} = 6/g^2_{run} - \frac{\partial (6/g_t^2)}{\partial \kappa_t}(\kappa_{run} - \kappa_t)$. The fit for $\kappa_t$ at $6/g^2 = 5.3, 5.1, 5.0$ and $4.9$ is $0.1579, 0.1713, 0.1772$ and $0.1827$ respectively.

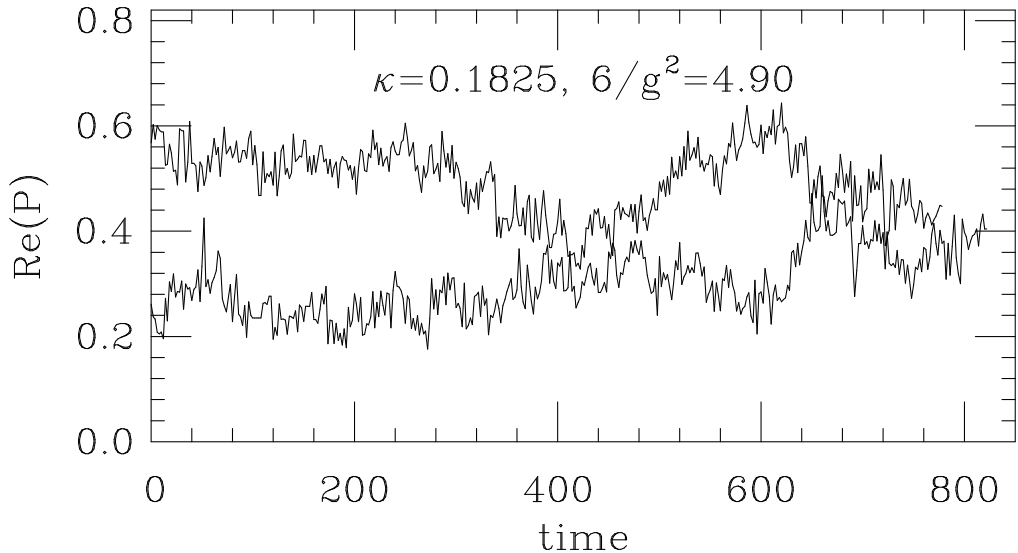

FIG. 4. Time history of the real part of the Polyakov loop for runs with hot and cold starts at $6/g^2 = 4.9$ and $\kappa = 0.1825$.

Figure 7 shows the average plaquette in these runs. Our normalization is such that the plaquette is three for a lattice of unit matrices. The plaquette also shows a sharp rise as the high temperature crossover is passed. Notice that for $6/g^2 = 5.3$ we have results for $N_t = 4$ and 6 showing that this increase is in fact due to the time size of the lattice, or the temperature.

The chiral condensate $\bar{\psi}\psi$ is less useful for Wilson quarks than for Kogut-Susskind quarks, since it does not go to zero in the high temperature phase without difficult subtractions. Nevertheless we plot it in Fig. 8. There is a clear drop in $\bar{\psi}\psi$ as the high temperature transition is crossed. This drop increases dramatically as $6/g^2$ decreases.

Perhaps the most physically relevant observable is the entropy. In Fig. 9 we plot $T \times s$ in units of $a^{-4}$. To give an idea of the normalization of this graph, for eight gluons in free field theory on an $8 \times 8 \times 20 \times 4$ lattice the gluon entropy would be $Ts_{glue,free} = 0.040 a^{-4}$, while for two flavors of free Wilson quarks at $\kappa = \kappa_c = 0.125$ the entropy would be $Ts_{quark,free} = 0.125 a^{-4}$. The effects of the lattice spacing and spatial size are very large here; in the continuum with infinite spatial extent these numbers are 0.027 and 0.036 respectively. Strangely, when we divide the entropy into gauge and fermion parts as in Eqs. 2 and 3 we find that the gauge entropy is comparable to the fermion entropy instead of much smaller as would be the case with free fields on a lattice of this size.

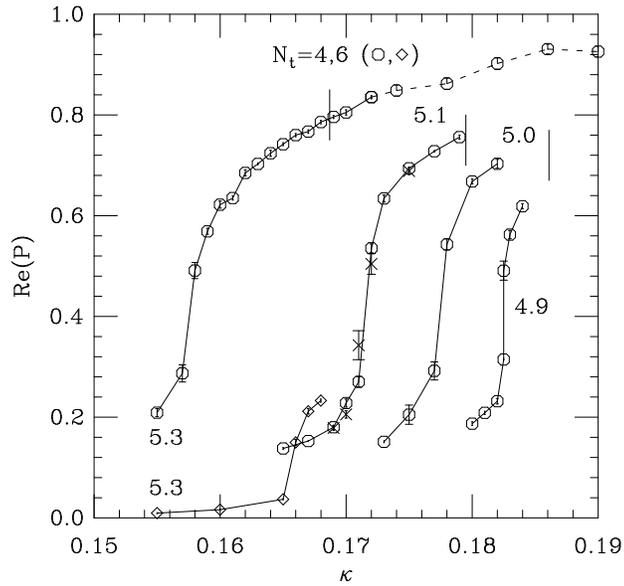

FIG. 5. Expectation value of the Polyakov loop as a function of $\kappa$ for the various values of $6/g^2$. Results are shown for $N_t = 4$ for $6/g^2 = 5.3$, 5.1, 5.0 and 4.9 (octagons). For $6/g^2 = 5.3$ we also show results for $N_t = 6$ (diamonds). The crosses along the $6/g^2 = 5.1$ line are results on a $6^2 \times 24 \times 4$ lattice at $6/g^2 = 5.1$, to show that the spatial size of the lattice is not greatly affecting the results. The dotted symbols extending the $6/g^2 = 5.3$ line are short runs on a $6^3 \times 4$ lattice, showing that the behavior is smooth out to very large $\kappa$. The vertical lines mark the zero temperature $\kappa_c$ for $6/g^2 = 5.2$, 5.1 and 5.0 respectively. (For $6/g^2 = 4.9$, $\kappa_c(T = 0) \approx 0.194$.)

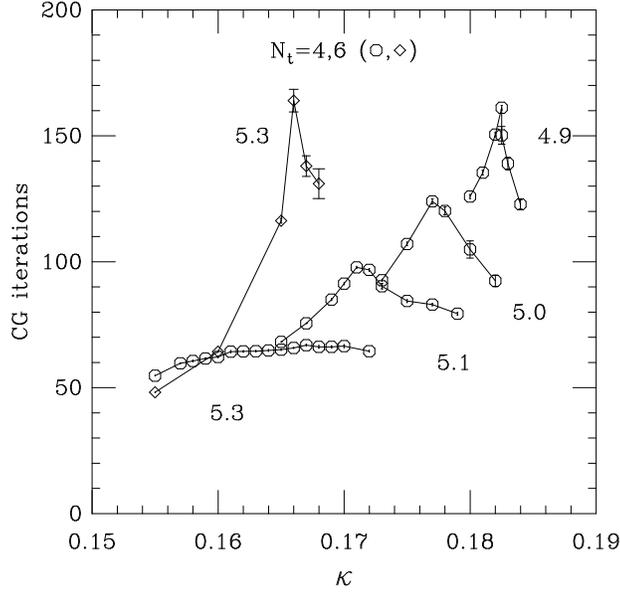

FIG. 6. Conjugate gradient iterations for updating step, as a function of $\kappa$ for the various values of $6/g^2$. Again the diamonds are $N_t = 6$ results at $6/g^2 = 5.3$.

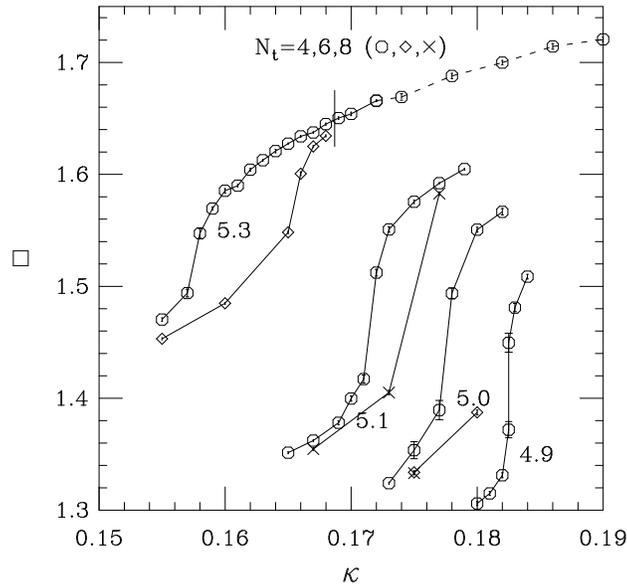

FIG. 7. Expectation value of the plaquette as a function of $\kappa$ for the various values of $6/g^2$. Here we included values for larger $N_t$ to emphasize the effect of the temperature. The dotted symbols for $6/g^2 = 5.3$ are short runs on a $6^3 \times 4$ lattice extending to $\kappa$ far beyond the zero temperature $\kappa_c$ shown by the vertical line.

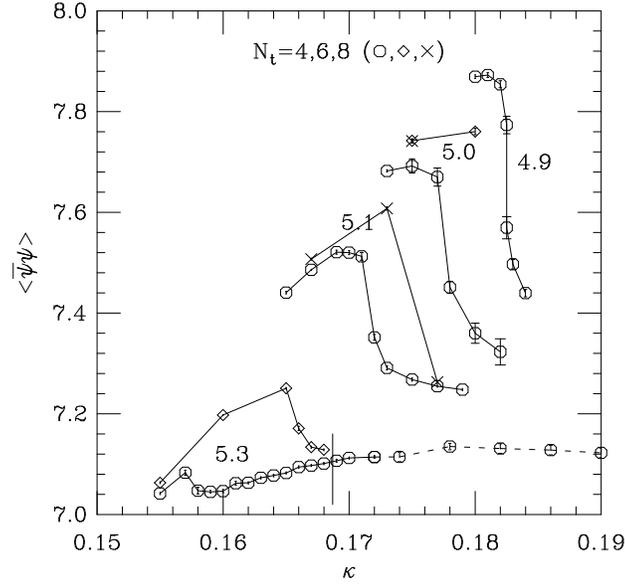

FIG. 8. Expectation value of $\bar{\psi}\psi$ as a function of $\kappa$ for the various values of $6/g^2$. The dotted symbols for $6/g^2 = 5.3$ are short runs on a $6^3 \times 4$ lattice extending to $\kappa$ far beyond the zero temperature $\kappa_c$ shown by the vertical line.

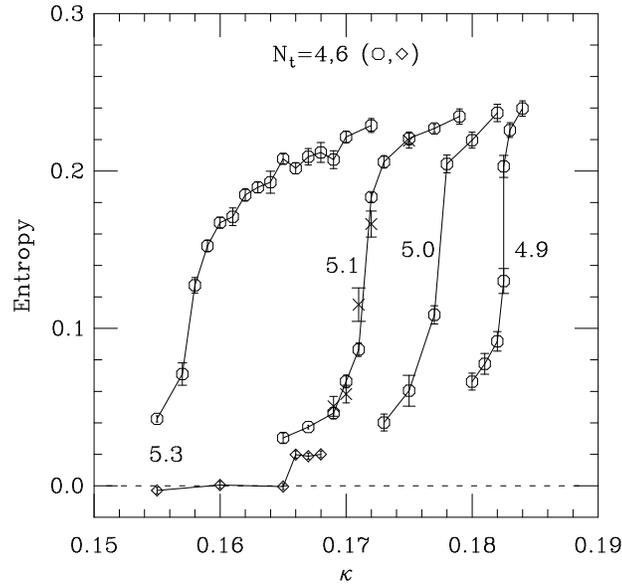

FIG. 9. Entropy (actually $T \times sa^4$) as a function of $\kappa$ for the various values of $6/g^2$. Again we show the $6^2 \times 24 \times 4$ results for comparison.

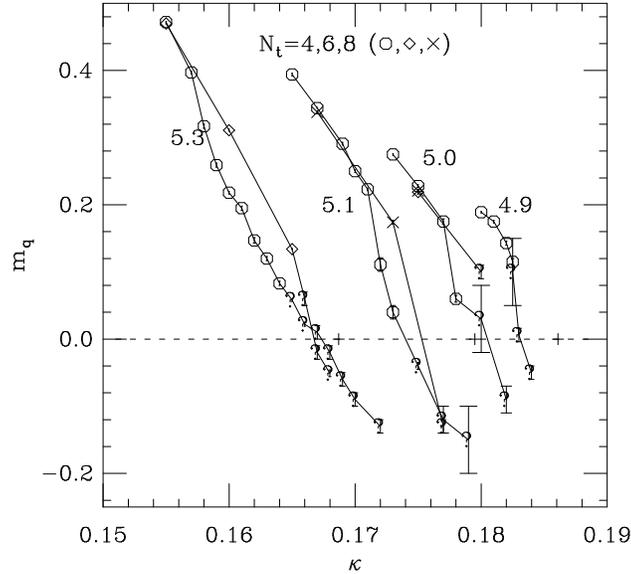

FIG. 10. Quark mass from the axial current as a function of $\kappa$ for the various values of $6/g^2$. Points marked with question marks indicate runs where we were unable to get consistent fits as a function of distance. The plus signs on the $m_q = 0$ line are the zero temperature $\kappa_c$ for $6/g^2 = 5.3$, 5.1 and 5.0.

The quark mass defined by the divergence of the axial pion propagator is plotted in Fig. 10. When this quark mass was small we had great difficulty in getting good fits to the forms in Eqs. 8 and 9. This is expected, because when the quark mass is small the amplitude for the propagator $A(z)$ is very small. Additionally, there is a tendency for the effective quark mass, or the quark mass coming from a fit over a short distance range, to increase with distance from the source. In cases where we were unable to get a fit with a satisfactory $\chi^2$ or where the quark mass was not convincingly independent of distance, we plot the point with a question mark in Fig. 10. To pursue this further we ran one of the difficult points, $6/g^2 = 5.3$ and $\kappa = 0.168$, on a $8^2 \times 40 \times 4$ lattice, allowing us to measure the ratio out to a distance of twenty. Figure 11 summarizes the results. In this figure we show the effective pion mass obtained from $PS(z)$ and $A(z)$ by fitting two two successive distances, and the quark mass obtained from simultaneously fitting both propagators at the two successive distances (a one degree of freedom fit). Unfortunately, in all other cases the lattice was only twenty sites long and we have to draw conclusions from distances less than ten. In Fig. 10 we see that when the $N_t = 4$ lattice enters the high temperature regime the pointlike axial current quark mass no longer agrees with the low temperature lattices ($N_t = 6$ and 8). The plusses at $m_q = 0$ in Fig. 10 are estimates for the zero temperature $\kappa_c$. The axial current quark masses go through zero at $\kappa$ less than the zero temperature $\kappa_c$. When the axial current quark mass vanishes, the system is in the high temperature phase for $\beta > 5.0$, while at $\beta = 4.9$ $k_t$ apears to coincide with the point where the axial current quark mass vanishes, within experimental uncertainty. Note however, that the pion screening mass in the confinement phase is still nonzero at the transition point at $\beta = 4.9$.

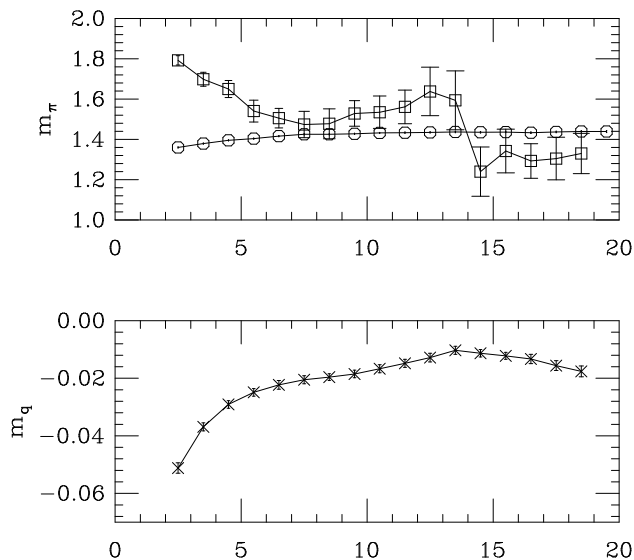

FIG. 11. Pion effective screening masses from $PS(z)$ (circles) and from $A(z)$ (squares), and the effective quark screening mass from their ratio. The results are from an $8^2 \times 40 \times 4$ lattice with $6/g^2 = 5.3$ and $\kappa = 0.168$.

In Fig. 12 we show the squared pion screening masses in these runs. Again we see an increasingly sharp dip at $\kappa_t$ as $6/g^2$ decreases and $\kappa$ increases. The appearance of the cusp at $\beta = 5.1$ coincides with the beginning of the region where the transition is abrupt. Screening masses for the $\pi$, $\rho$, $\sigma$ and $a_1$ mesons are shown in Figs. 13, 14 and 15. In all cases we see the screening masses coming together as the high temperature transition is crossed. However, we do not see any indication that the $\pi - \sigma$ or $\rho - a_1$ splittings in the high temperature regime are decreasing as $6/g^2$ decreases. Although the smaller pion masses in the cold regime suggest that chiral symmetry is being approached as we move toward smaller $6/g^2$ along the $\kappa_t$ line, we do not see this trend in the high temperature screening masses. Also notice that there are nonzero splittings between the parity partners at the points where the axial current quark mass is zero. Thus the vanishing of this quark mass is not an indicator for complete chiral symmetry restoration in the system.

To investigate the contributions of the doublers to thermodynamic quantities such as the entropy we measured the effective masses from the quark propagator in Landau gauge at a few values of $\kappa$ and $6/g^2$. We find that fitting the quark screening propagators is more difficult than fitting the meson propagators. In part this is because the quark propagators fluctuate more from configuration to configuration. There also seems to be a systematic trend toward larger effective quark masses at larger distances. With these caveats, the masses of the physical quark and the lightest doublers are given in Table IV. The fits were selected by choosing the largest fit range that gives an acceptable confidence level. The ranges and confidence levels are also given in Table IV.

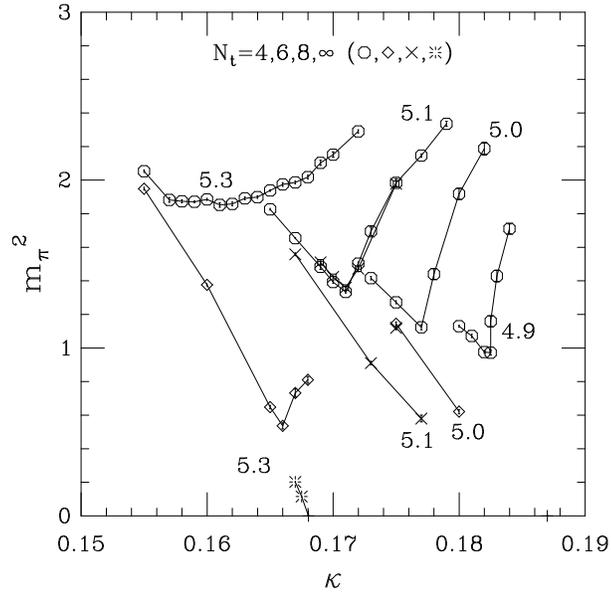

FIG. 12. Pion screening mass squared as a function of $\kappa$ for the various values of $6/g^2$. Again, the circles are for $N_t = 4$, the diamonds for $N_t = 6$ and the crosses for $N_t = 8$. The bursts are zero temperature pion masses from the HEMCGC collaboration at $6/g^2 = 5.3$.

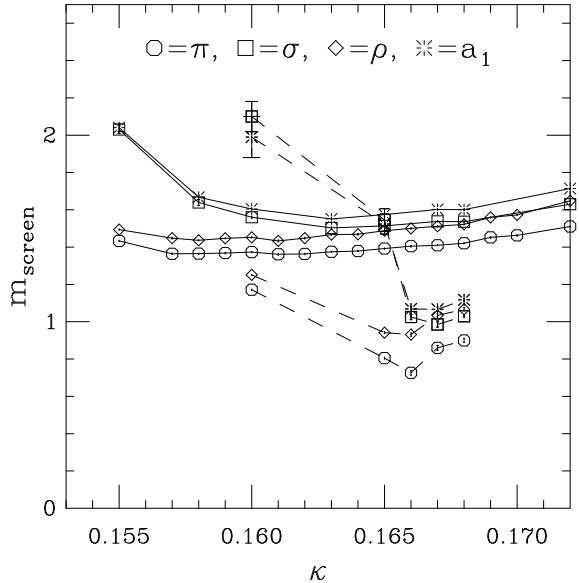

FIG. 13. Meson screening masses for $6/g^2 = 5.30$. The points connected by solid lines are for $N_t = 4$ and the points connected by dashed lines for $N_t = 6$.

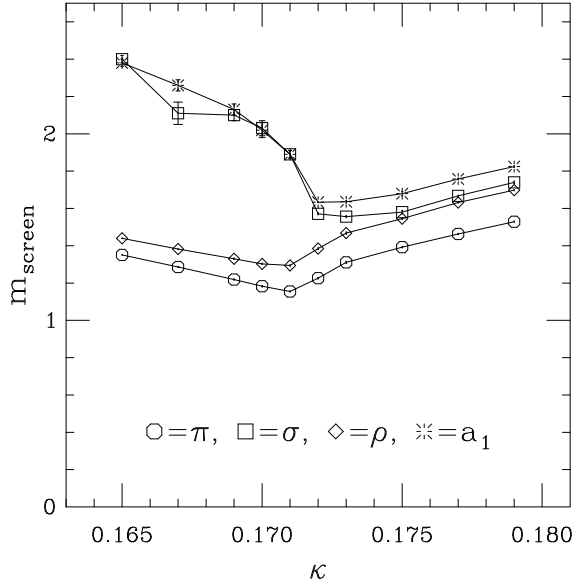

FIG. 14. Meson screening masses for $6/g^2 = 5.10$ at $N_t = 4$.

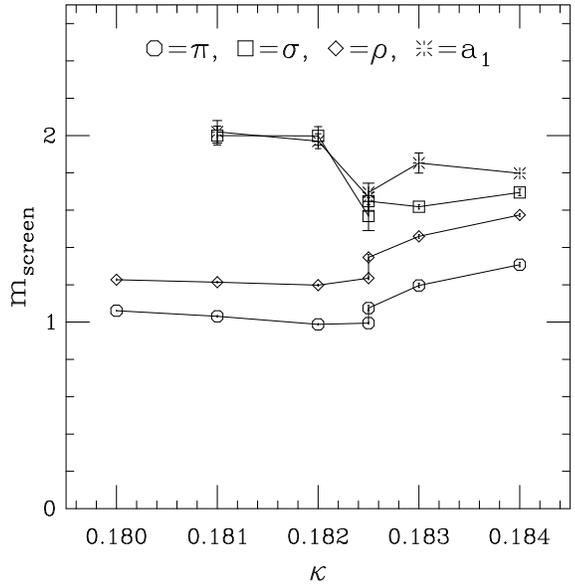

FIG. 15. Meson screening masses for $6/g^2 = 4.90$ at $N_t = 4$. The two points at $\kappa = 0.1825$ are from cold and hot starts.

| $N_t$ | $\kappa$ | $\beta$ | sign | $ma(0, \frac{\pi}{4})$ | $ma(\pi, \frac{\pi}{4})$ | $ma(0, \frac{3\pi}{4})$ | $\Delta ma_s$ | $\Delta ma_t$ | $q$ | range |
|---|---|---|---|---|---|---|---|---|---|---|
| 4 | 0.165 | 5.10 | + | 1.13(4) | 2.3(4) | 1.8(2) | 1.1(4) | 0.7(2) | 0.47 | 3-10 |
| 4 | 0.167 | 5.10 | + | 1.13(4) | 2.4(5) | 1.8(3) | 1.3(5) | 0.6(3) | 0.14 | 3-10 |
| 4 | 0.172 | 5.10 | + | 0.97(16) | 1.5(3) | 1.5(2) | 0.5(3) | 0.6(4) | 0.43 | 4-10 |
| 4 | 0.177 | 5.10 | − | 1.05(9) | 1.28(15) | 1.33(11) | 0.23(19) | 0.27(15) | 0.69 | 4-8 |
| 4 | 0.155 | 5.30 | + | 1.06(2) | 1.63(13) | 1.65(11) | 0.57(13) | 0.60(11) | 0.68 | 3-9 |
| 4 | 0.160 | 5.30 | + | 0.89(4) | 1.38(9) | 1.22(5) | 0.49(9) | 0.34(7) | 0.84 | 3-10 |
| 4 | 0.167 | 5.30 | − | 0.92(6) | 1.51(9) | 1.36(7) | 0.59(11) | 0.44(10) | 0.57 | 3-9 |

TABLE IV. The screening masses for the quark and the lightest doublers. $\Delta ma_{s(t)}$ is the difference of the spatial (temporal) doubler screening mass to the quark screening mass at the lowest momenta. The sign is for $G(k_t = \pi/4)$. The fits were done simultaneously to all three propagators taking into account the cross correlations. The confidence level $q$ and the range of each fit is also displayed.

## IV. CONCLUSIONS

The most naive expectation regarding the thermodynamics of two flavors of Wilson quarks at fixed $N_t$ is that there would be a line in the $\kappa, \beta$ plane at which a confinement-deconfinement transition occurs, that the transition would be smooth (crossover or second order), that the pion mass would smoothly decrease along that line, and that at some point, possibly corresponding to the point where the transition line crossed the zero temperature $\kappa_c - \beta$ line, the pion mass would go to zero. At that point one would have a finite temperature confinement-deconfinement or chirally restoring transition analogous to that seen in staggered fermions. Simple arguments [9] would put this point around $\beta = 5.0$ at $N_t = 4$.

These naive expectations are not borne out by the data. The chiral limit is reached at a very small $\beta$ value if it is reached at all. However, near $\beta = 5.0$ $N_t = 4$ Wilson thermodynamics displays a number of features which have no analogs in staggered fermion systems. The transition becomes very sharp, though not first order as far as we can tell. A cusp in the pion screening mass appears as one crosses from the confined to the deconfined phase. The axial vector quark mass becomes strongly $N_t$ dependent at this point and for small $N_t$ does not go to zero at its zero temperature $\kappa$ value (at fixed $\beta$). The sharp transition persists down to $\beta = 4.5, \kappa = 0.20$ or so, at which point it is once again becomes smooth. As far as we can tell, the zero temperature $\kappa = \kappa_c$ point plays no role in any $N_t = 4$ effects we have observed.

It is tempting to speculate that the crossover line in the $\kappa, 6/g^2$ plane is close to some phase boundary where the transition is steepest. We are currently exploring this region with $N_t = 6$, where preliminary results indicate a change in the nature of the high temperature transition around this value of $\kappa$.

Indicators for the nature of the high temperature phase give a somewhat mixed picture. It is clear from the meson screening masses and from $\bar{\psi}\psi$ that chiral symmetry is at least partially restored at high temperature. While the axial current quark mass goes to zero the $\pi - \sigma$ and $\rho - a_1$ splittings in the screening masses remain nonzero. Quark propagators in the Landau gauge suggest a large constituent quark mass at the transition, at least for $6/g^2 = 5.3$ and 5.1. This is consistent with earlier work [3] where at the $N_t = 4$ crossover point near these $(\beta, \kappa)$ values the zero temperature pion was found to be quite heavy.

Notice that the series of runs at $6/g^2 = 5.3$ extends to $\kappa$ significantly larger than the zero temperature $\kappa_c$, and there is no noticeable effect on any of the measured quantities when this $\kappa_c$ is crossed. (In fact, we have done short runs on $8^3 \times 4$ lattices for $\kappa$ as large as 0.19 at $6/g^2 = 5.3$ and seen no effects.) The sign of the propagator of the zero momentum quark, shown in table IV, is consistent with the sign of the axial current quark mass. Both of these quantities are behaving in the way one would expect in a free field theory at $\kappa > \kappa_c$.

## ACKNOWLEDGMENTS

These calculations were carried out on the iPSC/860, the Paragon and the nCUBE 2 at the San Diego Supercomputer Center, on the CM5 at the National Center for Supercomputing Applications, on 15 IBM/RS6000 workstations in the Physics Department at the University of Utah, on an IBM/RS6000 cluster at the Utah Supercomputing Institute and on our local workstations. We are grateful to the staffs of these centers for their help. We also thank Tony Anderson and Reshma Lal of Intel Scientific Computers for their help with the Paragon. We would like to thank Akira Ukawa and Frithjof Karsch for helpful discussions. Several of the authors have enjoyed the hospitality of the Institute for Nuclear Theory, the Institute for Theoretical Physics and the UCSB Physics department, where parts of this work were done. This research was supported in part by Department of Energy grants DE-2FG02–91ER–40628 , DE-AC02–84ER–40125, DE-AC02–86ER–40253, DE-FG02–85ER–40213, DE-FG03–90ER–40546, DE-FG02–91ER–40661, and National Science Foundation grants NSF–PHY90–08482, NSF–PHY93–09458, NSF–PHY91–16964, NSF–PHY89–04035 and NSF–PHY91–01853.

## APPENDIX

Expressions for the energy, pressure and $\bar{\psi}\psi$ are found by differentiating the partition function with respect to $1/T$, volume and quark mass, respectively. First, we write the action with adjustable lattice spacings in all directions. Introducing dimensionless parameters $\alpha_\mu$, we write the lattice spacing in the $\mu$ direction as $a_\mu = a\alpha_\mu$. Clearly this is redundant, since we have five parameters, $a$ and $\alpha_\mu$ to specify four lattice spacings, but it is convenient and symmetric. In the conventional notation of Karsch, $\xi = \alpha_i/\alpha_t$, where all the spatial $\alpha$'s are the same. When we are done taking derivatives, all the $\alpha_\mu$ will be set to one.

The partition function is

$$Z = \int [dU] e^{S_g + S_f} \tag{22}$$

where the gauge action is

$$S_g = \sum_x \sum_{\mu > \nu} \frac{2}{g_{\mu\nu}^2} \frac{\alpha_x \alpha_y \alpha_z \alpha_t}{\alpha_\mu^2 \alpha_\nu^2} \Box_{\mu\nu} \tag{23}$$

where $\Box_{\mu\nu}$ is the plaquette in the $\mu\nu$ plane normalized to three for unit matrices. We allow a different gauge coupling $g_{\mu\nu}$ in each plane. The fermion action is

$$S_f = \frac{n_f}{2} \operatorname{Tr} \log M^\dagger M \tag{24}$$

where

$$M = 1 - \kappa \sum_\mu \frac{\epsilon_\mu}{\alpha_\mu} \not{D}_\mu \tag{25}$$

where

$$\not{D}_\mu = (1 + \gamma_\mu) U_\mu(x) \delta_{y,x+\hat{\mu}} + (1 - \gamma_\mu) U_\mu^\dagger(x - \hat{\mu}) \delta_{y,x-\hat{\mu}} \tag{26}$$

The $\alpha_\mu$ in the coefficient of $\not{D}_\mu$ takes care of the dimensional scaling of the first derivative. Notice that we have made a somewhat arbitrary choice in $M$ when we scaled the irrelevant second derivative part with $\alpha_\mu$ in the same way that we scaled the first derivative part. The $\epsilon_\mu$ in the coefficient must be adjusted to get correlation functions to be Euclidean invariant. Its role is similar to the Karsch coefficents $C_\sigma$ and $C_\tau$ in the gauge action. Presumably $\epsilon_\mu$ has a power series expansion in $g$ just as $C_\sigma$ and $C_\tau$. Once again we have more parameters than we need: four $\epsilon_\mu$ and $\kappa$ for four directions. This parameterization is convenient because it includes the customary $\kappa$ and, later, $\kappa_c$ as parameters. We can fix the ambiguity well enough for our purposes by requiring that $\epsilon_\mu = 1$ when all the $\alpha_\mu$ are equal. In other words, if all directions are scaled by the same factor the only thing that changes is $\kappa$.

Let $\kappa_c$ be the value of $\kappa$ at which the pion mass and quark mass vanish, at least on an infinite lattice. Following free field theory, we introduce a quark mass

$$2ma = \kappa^{-1} - \kappa_c^{-1} \tag{27}$$

so that

$$M = \frac{1}{\kappa_c^{-1} + 2ma} \left( 2ma + \kappa_c^{-1} - \sum_\mu \frac{\epsilon_\mu}{\alpha_\mu} \not{D}_\mu \right) \tag{28}$$

Here $\kappa_c^{-1}$ will be a function of the couplings $g_{\mu\nu}$ and the scale factors $\alpha_\mu$. In free field theory, $\kappa_c^{-1} = 2 \sum_\mu \alpha_\mu^{-1}$. ($\epsilon_\mu = 1$ in free field theory.)

We find the energy, pressure and $\bar{\psi}\psi$ by differentiating the partition function:

$$\epsilon = -\frac{1}{V} \frac{\partial \log Z}{\partial \beta} \Big|_{V, m\ \text{constant}} \tag{29}$$

$$p = \frac{1}{\beta} \frac{\partial \log Z}{\partial V} \Big|_{\beta, m\ \text{constant}} \tag{30}$$

$$\bar{\psi}\psi = \frac{1}{\beta V} \frac{\partial \log Z}{\partial m} \Big|_{\beta, V\ \text{constant}} \tag{31}$$

Here $V$ is the volume, $V = a^3 \prod_i N_i \alpha_i = a^3 N_s^3 \alpha_s^3$. $\beta$ is the inverse temperature, $\beta = aN_t\alpha_t$. Here the energy and pressure derivatives are taken with $m$ constant, rather than with $\kappa$ constant. This is because $\kappa_c$ depends on the $\alpha_\mu$, so that if we distort the lattice while holding $\kappa$ fixed, the quark mass, and every physical mass, will vary sharply.

We change the temperature and volume using

$$\frac{\partial}{\partial \beta} = \frac{1}{N_t a} \frac{\partial}{\partial \alpha_t} \tag{32}$$

and

$$\frac{\partial}{\partial V} = \frac{1}{3N_s^3 a^3 \alpha_s^2} \frac{\partial}{\partial \alpha_s} \tag{33}$$

Alternatively, it may be easier to vary the volume by varying only one of the spatial lattice spacings

$$\frac{\partial}{\partial V} = \frac{1}{N_s^3 a^3 \alpha_x \alpha_y} \frac{\partial}{\partial \alpha_z} \tag{34}$$

The gauge energy and pressure are standard [18–20]. Following Ref. [18], we define two derivatives.

$$C_\sigma = -\frac{\partial g_{\mu\nu}^{-2}}{\partial \alpha_\lambda} \tag{35}$$

where $\lambda$ is not one of $\mu$ or $\nu$ and

$$C_\tau = -\frac{\partial g_{\mu\nu}^{-2}}{\partial \alpha_\lambda} \tag{36}$$

where $\lambda$ is one of $\mu$ or $\nu$.

Because stretching both the time and space directions is equivalent to changing the lattice spacing, $C_\sigma$ and $C_\tau$ are related to the beta function.

$$\left(\sum_\mu \frac{\partial}{\partial \alpha_\mu}\right) g_{ab}^{-2} = -2C_\sigma - 2C_\tau = \frac{\partial g^{-2}}{\partial \log(a)} \tag{37}$$

The contributions to the energy and pressure from $S_g$ and $S_f$ add. Doing the differentiation, and then setting the $\alpha_\mu$ to one, the gluon energy is

$$\epsilon_g a^4 = \frac{1}{N_s^3 N_t} \sum_x \left\langle \frac{6}{g^2} (\square_{st} - \square_{ss}) + 6C_\sigma \square_{ss} + 6C_\tau \square_{st} \right\rangle \tag{38}$$

Here $\square_{ss}$ and $\square_{st}$ are the space-space and space-time plaquettes, again normalized to three for a lattice of unit matrices.

For the gluon pressure, we find

$$p_g a^4 = \frac{1}{N_s^3 N_t} \sum_x \left\langle \frac{2}{g^2} (\square_{st} - \square_{ss}) - 2C_\sigma (\square_{ss} + 2\square_{st}) - 2C_\tau (2\square_{ss} + \square_{st}) \right\rangle \tag{39}$$

We also consider the linear combination $\epsilon + p$, the entropy.

$$s_g a^4 \frac{1}{aN_t} = \epsilon_g a^4 + p_g a^4$$
$$= \frac{1}{N_s^3 N_t} \sum_x \left\langle \left(\frac{8}{g^2} + 4(C_\tau - C_\sigma)\right)(\square_{st} - \square_{ss})\right\rangle \quad (40)$$

The entropy is obviously zero at $T = 0$.

Just as the gauge couplings vary with the lattice spacings, $\kappa_c^{-1}$ and $\epsilon_\mu$ vary with the lattice spacings as we try to hold $m$ fixed. There is an explicit dependence of $\kappa_c^{-1}$ on $m$ plus a dependence of $\kappa_c^{-1}$ on $g$, where $g$ is varying with the $\alpha_\mu$. Now it clearly doesn't matter which direction we stretch the lattice, since $\kappa_c^{-1}$ is defined on the infinite lattice, so

$$\frac{\partial \kappa_c^{-1}}{\partial \alpha_t} = \frac{\partial \kappa_c^{-1}}{\partial \alpha_x} \cdots . \quad (41)$$

Therefore

$$\frac{\partial \kappa_c^{-1}}{\partial \alpha_s}\bigg|_{\alpha_\mu=1} = 3 \frac{\partial \kappa_c^{-1}}{\partial \alpha_t}\bigg|_{\alpha_\mu=1} \quad (42)$$

There are two independent derivatives of the $\epsilon_\mu$, analogous to $C_\sigma$ and $C_\tau$. Define

$$B_\tau = \frac{\partial \epsilon_\mu}{\partial \alpha_\mu} \quad (43)$$

and

$$B_\sigma = \frac{\partial \epsilon_\mu}{\partial \alpha_\nu}, \quad \mu \neq \nu \quad (44)$$

To compute $\bar{\psi}\psi$ we can just set $\alpha_\mu$ and $\epsilon_\mu$ to one at the beginning.

$$a^3 \bar{\psi}\psi = \frac{1}{N_s^3 N_t} \frac{N_f}{2} \left\langle \frac{\partial}{\partial ma} \mathrm{Tr} \log M^\dagger M \right\rangle$$
$$= \frac{1}{N_s^3 N_t} \frac{N_f}{2} \left\langle \mathrm{Tr} \frac{1}{M^\dagger M} \left(\frac{\partial M^\dagger}{\partial ma} M + M^\dagger \frac{\partial M}{\partial ma}\right)\right\rangle$$
$$= \frac{1}{N_s^3 N_t} \frac{N_f}{2} \left\langle \mathrm{Tr} \left(\frac{1}{M^\dagger} \frac{\partial M^\dagger}{\partial ma} + \frac{1}{M} \frac{\partial M}{\partial ma}\right)\right\rangle \quad (45)$$

The two parts are complex conjugates, so keep only one and take twice the real part. Using Eq. (28)

$$a^3 \bar{\psi}\psi$$
$$= \frac{1}{N_s^3 N_t} \frac{N_f}{2} 2 \, \mathrm{Re} \left\langle \mathrm{Tr} \frac{1}{M^\dagger} \left(\frac{-2}{(\kappa_c^{-1} + 2ma)^2}\left(2ma + \kappa_c^{-1} - \sum_\mu \slashed{D}_\mu^\dagger\right) + 2\left(\frac{1}{\kappa_c^{-1} + 2ma}\right)\right)\right\rangle$$
$$= \frac{1}{N_s^3 N_t} \frac{4\kappa N_f}{2} \mathrm{Re} \left\langle \mathrm{Tr} \left(\frac{1}{M^\dagger} - 1\right)\right\rangle \quad (46)$$

where in the last step we used $1/\kappa_c^{-1} + 2ma = \kappa$. Looking at the derivation shows that the 1 in Eq. (46) comes from differentiating the $1/\kappa_c^{-1} + 2ma$ outside the parentheses in Eq. (28). Had we taken the fermion matrix to be

$$M = \left(2ma + \kappa_c^{-1} - \sum_\mu \frac{\epsilon_\mu}{\alpha_\mu} \not{D}_\mu\right) \tag{47}$$

this term would be absent. Since this latter form is closer to the usual continuum Lagrangian, we prefer

$$\bar{\psi}\psi = \frac{1}{N_s^3 N_t} \frac{4\kappa N_f}{2} \operatorname{Re}\left\langle \operatorname{Tr} \frac{1}{M^\dagger} \right\rangle \tag{48}$$

as our expression for $\bar{\psi}\psi$.

Now for the fermion energy, differentiate Eq. (25) and then set $\alpha_\mu$ and $\epsilon_\mu$ to one:

$$\begin{aligned}
\epsilon_f a^4 &= \frac{-1}{N_s^3 N_t} \frac{N_f}{2} \left\langle 2\operatorname{Re}\operatorname{Tr} \frac{1}{M^\dagger} \frac{\partial M^\dagger}{\partial \alpha_t} \right\rangle \\
&= \frac{-2}{N_s^3 N_t} \frac{N_f}{2} \operatorname{Re}\left\langle \frac{1}{M^\dagger} \left[\frac{-1}{(\kappa_c^{-1} + 2ma)^2}\left(2ma + \kappa_c^{-1} - \sum_\mu \not{D}_\mu^\dagger\right)\frac{\partial \kappa_c^{-1}}{\partial \alpha_t}\right.\right. \\
&\quad \left.\left. +\frac{1}{\kappa_c^{-1} + 2ma}\left(\not{D}_0^\dagger - \sum_\mu \frac{\partial \epsilon_\mu}{\partial \alpha_t} \not{D}_\mu^\dagger + \frac{\partial \kappa_c^{-1}}{\partial \alpha_t}\right)\right]\right\rangle \\
&= \frac{2}{N_s^3 N_t} \frac{\kappa N_f}{2} \operatorname{Re}\left\langle \frac{\partial \kappa_c^{-1}}{\partial \alpha_t}\left(1 - \frac{1}{M^\dagger}\right) - \frac{1}{M^\dagger}\left(\not{D}_0^\dagger - B_\tau \not{D}_0^\dagger - \sum_i B_\sigma \not{D}_i^\dagger\right) \right\rangle
\end{aligned} \tag{49,50}$$

where in the last step we used $1/\kappa_c^{-1} + 2ma = \kappa$.

For the fermion pressure:

$$\begin{aligned}
p_f a^4 &= \frac{1}{3N_s^3 N_t} \frac{N_f}{2} \left\langle 2\operatorname{Re}\operatorname{Tr} \frac{1}{M^\dagger} \frac{\partial M^\dagger}{\partial \alpha_s} \right\rangle \\
&= \frac{2}{3N_s^3 N_t} \frac{N_f}{2} \operatorname{Re}\left\langle \frac{1}{M^\dagger} \left[\frac{-1}{(\kappa_c^{-1} + 2ma)^2}\left(2ma + \kappa_c^{-1} - \sum_\mu \not{D}_\mu^\dagger\right)\frac{\partial \kappa_c^{-1}}{\partial \alpha_s}\right.\right. \\
&\quad \left.\left. +\frac{1}{\kappa_c^{-1} + 2ma}\left(\sum_i \not{D}_i - \sum_\mu \frac{\partial \epsilon_\mu}{\partial \alpha_s} \not{D}_\mu^\dagger + \frac{\partial \kappa_c^{-1}}{\partial \alpha_s}\right)\right]\right\rangle \\
&= \frac{-2}{3N_s^3 N_t} \frac{\kappa N_f}{2} \operatorname{Re}\left\langle \frac{\partial \kappa_c^{-1}}{\partial \alpha_s}\left(1 - \frac{1}{M^\dagger}\right) + \frac{1}{M^\dagger}\left(3 B_\sigma \not{D}_0^\dagger + (2B_\sigma + B_\tau)\sum_i \not{D}_i^\dagger - \sum_i \not{D}_i^\dagger\right) \right\rangle
\end{aligned} \tag{51,52}$$

Just as for $\bar{\psi}\psi$, the 1 term in $\epsilon_f$ and $p_f$ comes from differentiating the overall factor of $\kappa^{-1}$. It can be included or not, as desired. It will cancel when the finite parts of the energy and pressure are calculated by subtracting the zero temperature result from the nonzero temperature result. However, the $\frac{1}{M^\dagger}\frac{\partial \kappa_c^{-1}}{\partial \alpha_t}$ term will not cancel out, since $\bar{\psi}\psi$ is temperature dependent. In practice $\langle \operatorname{Tr} 1/M^\dagger \rangle$ is fairly close to $\langle \operatorname{Tr} 1 \rangle = 4 \times 3$, so numerically it may be best to leave the 1 in. Then we would use

$$1 - \frac{1}{M^\dagger} = \frac{\kappa \rlap{/}{D}^\dagger}{M^\dagger} \tag{53}$$

to express the energy and pressure just in terms of the expectation values of the spatial and temporal components of $\rlap{/}{D}^\dagger$.

Much of the difficulty cancels out if we look at the entropy, or sum of energy and pressure:

$$\begin{aligned} s_f a^4 \frac{1}{aN_t} &= \epsilon_f a^4 + p_f a^4 \\ &= \frac{-2}{N_s^3 N_t} \frac{\kappa N_f}{2} \operatorname{Re} \left\langle \operatorname{Tr} \frac{1}{M^\dagger} (1 + (B_\tau - B_\sigma)) \left( \rlap{/}{D}_0 - \frac{1}{3} \sum_i \rlap{/}{D}_i \right) \right\rangle \end{aligned} \tag{54}$$

The relation

$$\frac{\partial \kappa_c^{-1}}{\partial \alpha_s} = 3 \frac{\partial \kappa_c^{-1}}{\partial \alpha_t} \tag{55}$$

resulted in all the derivative terms cancelling. (Remember, by $\partial/\partial \alpha_s$ we mean $\partial/\partial \alpha_x + \partial/\partial \alpha_y + \partial/\partial \alpha_z$ — vary all the spatial $\alpha_i$ together.) As usual, the entropy is obviously zero at $T = 0$, where $\rlap{/}{D}_i = \rlap{/}{D}_0$. Since no zero temperature subtraction is required for the entropy, the terms involving $B_\tau$ and $B_\sigma$ will be higher order in $g^2$ than the "1" term, and we have neglected them in Eq. (3).

Obviously, the big problem in getting the energy and pressure separately is to find $\frac{\partial \kappa_c^{-1}}{\partial \alpha_\mu}$ and $B_\tau$ and $B_\sigma$. From Euclidean invariance, $\frac{\partial \kappa_c^{-1}}{\partial \alpha_\mu}$ is independent of $\mu$. The variable $\kappa_c^{-1}$ depends on the $\alpha_\mu$ in two ways. First, there is an "explicit" dependence. From examining the fermion matrix, Eq. (25), we see that if all the $\alpha_\mu$ are scaled together with $g$ held fixed, $\kappa_c$ is proportional to $\alpha$. Thus

$$\left. \frac{\partial \kappa_c^{-1}}{\partial \alpha_\mu} \right|_{explicit} = \frac{-\kappa_c^{-1}}{4} \tag{56}$$

Secondly, there is an "implicit" dependence of $\kappa_c^{-1}$ on $\alpha_\mu$ coming from the fact that $\kappa_c$ depends on $g^{-2}$, and we adjust the $g_{\mu\nu}^{-2}$ as we adjust the $\alpha_\mu$. Again, Euclidean invariance says

$$\left. \frac{\partial \kappa_c^{-1}}{\partial \alpha_\mu} \right|_{implicit} = \frac{1}{4} \frac{\partial \kappa_c^{-1}}{\partial \log(a)} , \tag{57}$$

so only the beta function appears.

$$\begin{aligned} \left. \frac{\partial \kappa_c^{-1}}{\partial \alpha_\mu} \right|_{implicit} &= \frac{1}{4} \frac{\partial \kappa_c^{-1}}{\partial \log(a)} \\ &= \frac{1}{4} \frac{\partial \kappa_c^{-1}}{\partial 6/g^2} \frac{\partial 6/g^2}{\partial \log(a)} \end{aligned} \tag{58}$$

We could proceed by estimating $\partial \kappa_c^{-1}/\partial 6/g^2$ from our data at various values of $6/g^2$, or from correlations of the hadron propagators with the plaquette. Similarly, we could take a beta function either from perturbation theory or from some set of lattice simulations. We will not solve this problem here, so we will only quote the entropy rather than the energy and pressure separately.